\documentclass[prd,amsmath,amssymb,floatfix]{revtex4}

\usepackage{graphicx}

\begin{document}
                                                                                
\title{Asymmetric Spatiotemporal
Evolution of Prebiotic Homochirality} 
\author{Marcelo Gleiser}
\email{gleiser@dartmouth.edu}

\affiliation{Department of Physics and Astronomy, Dartmouth College
Hanover, NH 03755, USA}

\begin{abstract}
The role of asymmetry on the evolution of prebiotic homochirality is
investigated in the context of autocatalytic polymerization reaction networks.
A model featuring enantiometric cross-inhibition and chiral bias is used to
study the diffusion equations controlling the spatiotemporal development of
left and right-handed domains. Bounds on the chiral bias are obtained
based on present-day constraints on the emergence of life on early
Earth. The viability of biasing mechanisms such as weak neutral currents and
circularly polarized UV light is discussed. The results can be applied to any
hypothetical planetary platform.
\end{abstract}
\keywords{prebiotic chemistry, origin of life,
early planetary environments}
    
\maketitle

\section{Introduction}

The emergence of biomolecular homochirality in prebiotic Earth is a crucial
step in the early history of life \cite{AG93, Bonner}. It is well-known that
chiral selectivity plays a key role in the biochemistry of living systems:
amino acids in proteins are left-handed while sugars are right-handed. 
However, laboratory syntheses produce racemic results. This is somewhat 
surprising, given that statistical fluctuations of reactants will invariably
bias one enantiometer over the other \cite{Blackmond04}:  even though every
synthesis is {\it ab initio} asymmetric \cite{Dunitz}, the enantiometric excess
is nevertheless erased as the reactions unfold. An important exception is the
reaction by Soai and  coworkers, where a small initial enantiometric excess is 
effectively amplified in the autocatalytic alkylation of pyrimidyl  aldehydes
with dialkylzincs \cite{Soai}. As stressed by Blackmond \cite{Blackmond04}, 
Soai's reaction succeeds because it features the needed autocatalytic
behavior proposed originally by Frank \cite{Frank53} with enantiometric
cross-inhibition catalysed by dimers.

It is unlikely that the specific chemistry of the Soai reaction occurred in
early-Earth. However, it displays the relevant signatures of a realistic
homochirality-inducing reaction network: autocatalysis, enantiometric
cross-inhibition, and enzymatic enhancement performed by dimers or by larger
chirally-pure chains.  In the present work, we will investigate the
spatiotemporal dynamics of a reaction network recently proposed by Sandars
which shares these features  \cite{Sandars03}. To it, we will add an explicit
chiral bias, in order to investigate the efficacy of intrinsic and extrinsic
biasing mechanisms proposed in the literature. 
A first step in this direction can be found in the work by Brandenburg
{\it et al.}, who studied an extension of Sandar's model including bias but no
spatial dependence \cite{BAHN}.
By intrinsic we mean either
biasing effects related to fundamental physics, such as parity-violating weak
neutral currents (WNC) \cite{WNC0,KN83} or -- given that we know little of
early-Earth's prebiotic chemistry and even less of other possible life-bearing
planetary platforms  \cite{Orgel98} --  to some as yet unknown chemical process.
By extrinsic we mean possible environmental influences, such as
circularly-polarized UV light (CPL) from, for example, active star-formation
regions \cite{Lucas05} or direct seeding of chiral compounds by meteoritic
bombardment \cite{Engel97, Cronin98}. The spatiotemporal dynamics of the
reaction network will be shown to be equivalent to a two-phase system
undergoing a symmetry-breaking phase transition characterized by the formation
of competing domains of opposite chirality. The evolution of the domain network
is sensitive to Earth's early environmental history and to the magnitude of the
chiral bias.  Using the time-scale associated with the emergence of life on
Earth it is possible to obtain a lower bound on the bias.  In particular, it
will be shown that the very small bias from WNC is inefficient to generate 
homochiral conditions. For CPL the situation is less clear due to  uncertainty
in the nature and duration of sources, but still highly unlikely. The formalism
is set up to be applicable to any planetary platform.

The paper is organized as follows: In section 2 we introduce the biased 
polymerization model and its pertinent rate equations. In section 3 we describe
the reduced ($n=2$) model and how the net chirality can be interpreted as a 
continuous order parameter satisfying an effective potential. This allows us to
introduce explicitly spatial dependence in the study of biased polymerization.
In section 4 we describe the dynamics of homochirality using techniques from
the theory of phase transitions with mean-field Ginzburg-Landau models in the
limit of no bias. In section 5 we generalize our results to include a small
bias, describing in detail the wall dynamics in this case and the time-scales
associated with the development of homochirality in early-Earth. We also
investigate if the onset of prebiotic homochirality on early-Earth could have
been the result of a nucleation event.We conclude in section 6 with a summary
of our results and an outlook to future work.

\section{Modeling Biased Polymerization}

Sandar's model describes how long chains of homochiral polymers may evolve
from a gradual build up of chiral monomers \cite{Sandars03}.  In order to reach
homochirality two processes are needed: reactions must be autocatalytic so that
longer, chirally-pure chains may be synthesized. In addition, a mechanism for
chiral amplification is also needed. This amplification may be achieved in a
number of ways \cite{KN83,AG93,Plasson04}. Sandars included  both enantiometric
cross-inhibition and an enzymatic enhancement catalysed by the longest chain in
the reactor pool.

Consider a polymer with $n$ left-handed monomers, $L_n$. It may grow by 
aggregating a left-handed monomer $L_1$ or it may instead be inhibited by the
addition of a right-handed monomer $R_1$ to either of its ends. Writing the
reaction rates as $k_S$ and  $k_I$, the reaction network can be written as
\cite{Sandars03}:
\begin{eqnarray}
\label{sandarseqs}
L_n + L_1 &\stackrel{2k_S}{\longrightarrow}& L_{n+1} \nonumber \\ 
L_n + R_1 &\stackrel{2k_I}{\longrightarrow}& L_nR_1 \nonumber \\
L_1 + L_nR_1 &\stackrel{k_S}{\longrightarrow}& L_{n+1}R_1 \nonumber \\
R_1 + L_nR_1 &\stackrel{k_I}{\longrightarrow}& R_1L_nR_1~,
\end{eqnarray}
supplemented by the four opposite reactions for right-handed polymers by
interchanging $L\leftrightarrow R$. The network includes a substrate $S$ from
where both left and right-handed monomers are generated. The rate at which
monomers are generated may depend on several factors. It may be due to already
existing polymers with an enzymatic enhancement denoted here by $C_{L(R)}$ for
left(right)-handed monomers. Sandars wrote $C_L = L_N$ and $C_R
= R_N$, where $N$ is the largest polymer in the substrate.  If $N=2$, the case
we investigate here, we can model the catalytic role of dimers 
\cite{Blackmond04}. Wattis and Coveney \cite{WC05} proposed instead
$C_L = \sum L_n$ and $C_R = \sum R_n$, while Brandenburg {\it et al.}
\cite{BAHN} suggested a weighted sum, $C_L=\sum nL_n$ and $C_R=\sum nR_n$.
Motivated by mathematical simplicity and by Soai's reactions, we will
follow Sandars.

Another factor that may influence the production rate of monomers is an
explicit bias towards a specific handedness. We assign a chiral-specific
reaction rate $k_{L(R)}$ such that the generation of left and right-handed
monomers from the substrate $S$ is written as 
\begin{eqnarray}
\label{substrate}
S &\stackrel{k_L}{\longrightarrow}& L_1 \nonumber \\
S &\stackrel{k_R}{\longrightarrow}& R_1.
\end{eqnarray}
The reaction rates are
related to the equilibrium population of  each handedness as $k_L\propto
\exp[-E_L/k_BT]$ and  $k_R\propto \exp[-(E_L+E_f)/k_BT]$, where $k_B$ is
Boltzmann's constant, $T$ is the temperature, and $E_f$ denotes the  energy
difference between the two enantiometers, here chosen arbitrarily to suppress 
the right-handed monomers. Note that similar results would have been obtained
by considering $E_{L(R)}$ to be the activation energy for forming
$L(R)$ molecules, and $E_f$ the difference in activation energy between
enantiometers.

Kondepudi and Nelson \cite{KN85} used a similar parameterization to express the
bias due to parity violation in the weak nuclear interactions, which has been
estimated to be $g\equiv E_f/k_BT \sim 10^{-17-18}$ at room temperature
\cite{Salam91,WNC3,WNC4}. On the other hand, CPL biasing depends on a number of
unknowns such as the nature of the UV source, its distance and duration and
it's harder to estimate  \cite{CPL2,KN85}. Since in general $g\ll 1$, one
obtains $k_L/k_R\simeq 1+g$. 
Introducing the average reaction rate $k_C\equiv \frac{k_L + k_R}{2}$, we
can express the left and right-handed reaction rates as
\begin{eqnarray}
\label{biased_rates}
k_L &=& k_C(1+g/2) + {\cal O}(g^2) \nonumber \\
k_R &=& k_C(1-g/2) + {\cal O}(g^2)~.
\end{eqnarray}

If the concentration of the substrate $[S]$
is maintained by a source $Q$, it will obey the equation,
\begin{equation}
\label{substrateconc}
\frac{d[S]}{dt} = Q - (Q_L + Q_R)~,
\end{equation}
where $Q_{L(R)}$ are the sources for left(right)-handed
monomers. From eqs. \ref{substrate} and \ref{biased_rates},
the sources can be written as
\begin{eqnarray}
\label{lrsources}
Q_L &=& k_C(1+g/2)[S](pC_L+qC_R) \nonumber \\
Q_R &=& k_C(1-g/2)[S](pC_R+qC_L)~,
\end{eqnarray}
where we introduced the fidelity $f$ of the enzymatic reactions, written in
terms of $p$ and $q$ as $p\equiv (1+f)/2$ and $q\equiv (1-f)/2$. In the absence
of bias ($g=0$), the fidelity $f$ controls 
the evolution of the net enantiometric excess.

As has been thoroughly discussed in the literature 
\cite{K96,Sandars03,WC05,BAHN}, the dynamical system defined by the
polymerization equations for a given value of $N$ shows a bifurcation behavior
at a certain critical value of $f$, $f_c$. The specific value of $f_c$ depends
on the choice made for the enzymatic enhancements $C_{L(R)}$ and on the ratio
of reaction rates $k_I/k_S$, but the behavior is qualitatively the same.
Brandenburg {\it et al.}, with $k_I/k_S=1$, obtained $f_c\simeq 0.38$, with
$f_c$ increasing with weaker cross-inhibition. In the limit $k_I\rightarrow 0$,
$f_c\rightarrow 1$ and no enantiometric excess develops \cite{BAHN}.

The full set of reaction rate equations governing the
behavior of an $n$-polymer ($n=1,\dots, N$) system consists of the following
equations \cite{Sandars03,BAHN}: the equation for the substrate concentration, 
eq. \ref{substrateconc}; the equations for the left and right monomers,
\begin{eqnarray}
\frac{d[L_1]}{dt} &=& Q_L-\lambda_L[L_1]~,\nonumber\\
\frac{d[R_1]}{dt} &=& Q_R-\lambda_R[R_1]~,
\end{eqnarray}
where
\begin{eqnarray}
\lambda_L&=&2k_S\sum_{n=1}^{N-1}[L_n]+2k_I\sum_{n=1}^{N-1}[R_n] \nonumber \\
&+&k_S\sum_{n=2}^{N-1}[L_nR_1]+k_I\sum_{n=2}^{N-1}[R_nL_1]~,
\end{eqnarray}
and
\begin{eqnarray}
\lambda_R&=&2k_S\sum_{n=1}^{N-1}[R_n]+2k_I\sum_{n=1}^{N-1}[L_n] \nonumber \\
&+&k_S\sum_{n=2}^{N-1}[R_nL_1]+k_I\sum_{n=2}^{N-1}[L_nR_1]~;
\end{eqnarray}
and, finally, the rate eqns. for $n\geq 2$,
\begin{equation}
\label{rateeqs}
\frac{d[L_n]}{dt}=2k_S[L_1]\left ([L_{n-1}]-[L_n]\right )-2k_I[L_n][R_1]~,
\end{equation}
supplemented by the ones obtained substituting $L\rightarrow R$.
The factors of $2$ on the rhs reflect that monomers may attach to either
end of the chain. For $n=2$, however, one must discount this for the interaction
of 2 single monomers.

\section{Reduced Biased Model}

Previous authors \cite{Sandars03, BAHN, WC05} have explored the evolution of
the reaction network for different values of $N$. Here, we are interested in
investigating not only the temporal evolution of the various concentrations
($[L_n],~[R_n]$) but also their spatial behavior in the presence of bias.  We
are motivated by interesting work by Saito and Huyga \cite{SH04}, who
investigated spatial proliferation of left and right-handed polymers in the
context of equilibrium Monte-Carlo methods and, in particular, by that of
Brandenburg and Multam\"aki \cite{BM} (henceforth BM),  where the
spatiotemporal evolution of left and right-handed reaction networks was
investigated in the absence of chiral bias.

\subsection{Biased Polymerization Equations} 

As remarked by Gayathri and Rao \cite{GR05}, taking the concentrations to be
functions of position implies that the number of molecules per unit volume is
assumed to be large enough so that the concentrations  vary smoothly with space
and time. In other words, the  concentrations are defined in a coarse-grained
volume which, of course, must be larger than the smallest relevant
distance-scale ($\xi$), to be derived below. The chemical mixture is then
defined in block-volumes which are multiples of $\sim\xi^3$. This is
essentially the procedure adopted in studying mean-field models of phase
transitions in the Ising universality class as, for example, in ferromagnetic
phase transitions, where the order parameter is the coarse-grained
magnetization over a block of spins \cite{Langerrev}. Indeed, in a recent work
\cite{GT} Gleiser and Thorarinson (GT)  demonstrated that chiral symmetry
breaking in the context of the continuous model of BM can be understood in
terms of a second-order phase transition with a critical ``temperature''
determined by the strength of the coupling between the reaction network and the
external environment. One cannot speak consistently of symmetry breaking
without including spatial dependence.

Adding spatial dependence to the reaction network greatly complicates its
study, as we must investigate coupled PDEs as opposed to ODEs. Fortunately, as
remarked by BM, it is possible to truncate the system to $N=2$ and still
capture its essential behavior, the dynamics leading (or not) to homochirality
within a large volume ${\cal V} \gg \xi^3$. Given the catalytic role of dimers
in the Soai reaction \cite{Blackmond04} we will investigate the biased
spatiotemporal dynamics of reaction networks with $N=2$.

The great practical advantage of the truncation is that it elegantly reduces
the system to an effective scalar field theory, where the field -- the order
parameter -- determines the net chirality in a given volume \cite{BM}.

The reaction network is further simplified by assuming that the rate of change
of $[L_2]$ and $[R_2]$ is much slower than that of $[L_1]$ and $[R_1]$. The
same for the substrate $[S]$, so that $d[S]/dt = Q-(Q_L+Q_R) \simeq 0$. This
approximation is known as the adiabatic elimination of rapidly adjusting
variables \cite{Haken83}: the typical time-scale for changes in the
concentrations of larger chains such as $[L_2]$
and $[R_2]$, which depend on (or are enslaved by) the concentrations of $[L_1]$
and $[R_1]$, are slower than those for the concentrations $[L_1]$ and $[R_1]$
themselves. This approximation breaks down in the unlikely situation where
the syntheses of dimers and higher chains have similar time-scales as those 
for monomers.

Using the adiabatic approximation in
eq. \ref{rateeqs} for $n=2$,
we can express the concentrations $[L_2]$ and $[R_2]$ in terms of the 
concentrations for the monomers as
\begin{eqnarray}
\label{adiabapprox}
[L_2] = \frac{[L_1]^2}{2[L_1]+2k_S/k_I[R_1]};
[R_2] = \frac{[R_1]^2}{2[R_1]+2k_S/k_I[L_1]}~.
\end{eqnarray}
Also, using eq. \ref{substrateconc} with $C_L=[L_2]$ and $C_R=[R_2]$, 
the equations for the monomers can be written as
\begin{eqnarray}
\label{monomereqs}
\frac{d[L_1]}{dt} &=& Q\frac{\left (1+g/2\right )
\left (p[L_2]+q[R_2]\right )}{\left ([L_2]+[R_2]\right )
\left [1 +\frac{g}{2}f\left
(\frac{[L_2]-[R_2]}{[L_2]+[R_2]}\right )\right ]} - \lambda_L[L_1] \nonumber \\
\frac{d[R_1]}{dt} &=& Q\frac{\left (1-g/2\right )
\left (p[R_2]+q[L_2]\right )}{\left ([L_2]+[R_2]\right )
\left [1 +\frac{g}{2}f\left
(\frac{[L_2]-[R_2]}{[L_2]+[R_2]}\right )\right ]} - \lambda_R[R_1]~.
\end{eqnarray}
Since $g\ll 1$, we can expand the rhs of eqs. \ref{monomereqs}
and greatly simplify the equations for the monomers. Once we substitute eqns.
\ref{adiabapprox} into eqns. \ref{monomereqs} we are left with two equations
for the two unknowns, $[L_1]$ and $[R_1]$.

It is useful at this point to introduce the dimensionless symmetric and
asymmetric variables, ${\cal S}\equiv X+Y$ and ${\cal A}\equiv X-Y$,
where $X\equiv [L_1](2k_S/Q)^{1/2}$, and $Y\equiv [R_1](2k_S/Q)^{1/2}$  are
dimensionless concentrations. We can thus add and subtract the equations for the
monomers in order to obtain the equations satisfied by the variables
${\cal S}$ and ${\cal A}$, respectively. After some algebra we get,
for $k_S/k_I=1$, the biased polymerization equations,
\begin{eqnarray}
\label{SAeqs1}
\lambda_0^{-1}\frac{d{\cal S}}{dt}&=&1-S^2 \nonumber \\
\lambda_0^{-1}\frac{d{\cal A}}{dt}&=&2f\frac{{\cal S}{\cal A}}{{\cal S}^2+
{\cal A}^2} - {\cal S}{\cal A} + \frac{g}{2}\left [1 - 4f^2\left (
	\frac{{\cal S}
{\cal A}}{{\cal S}^2+{\cal A}^2}\right )^2\right ]~,
\end{eqnarray}
where the parameter $\lambda_0\equiv (2k_SQ)^{1/2}$, has dimension of inverse
time.  ${\cal S}=1$ is a fixed point: the system will tend towards this value
at time-scales of order $\lambda_0^{-1}$, independently of $g$.  
With ${\cal S}=1$
and $g=0$, the equation for the chiral asymmetry has fixed points at ${\cal
A}=0,~\pm\sqrt{2f-1}$.  An enantiometric excess is only possible for
$f>f_c=1/2$.

Setting ${\cal S}=1$, the equation for ${\cal A}$ can be written as
$\lambda_0^{-1}\dot{\cal A} = -\partial V/\partial {\cal A}$, where the
dot denotes time derivative. The ``potential'' $V$ controlling the
evolution of ${\cal A}$ is an asymmetric double-well, (symmetric for $g=0$)
\begin{equation}
\label{pot}
V({\cal A}) = \frac{{\cal A}^2}{2}-f\ln(1+{\cal A}^2)-
\frac{g}{2}{\cal A}-gf^2\left [\frac{{\cal A}}{1+{\cal A}^2}-\arctan({\cal A})
\right ]~.
\end{equation}

In figure 1 we show the potential $V({\cal A})$ for various values of the
fidelity $f$ and the asymmetry $g$. Note that complete chiral separation occurs
only for $f=1$ and is forbidden for $f\geq 1/2$ (left). The presence of
asymmetry ($g\neq 0$, right) clearly biases one of the chiralities, leaving the
minima at ${\cal A}=\pm 1$ unchanged for $f=1$. 

\begin{figure}
 \centering
 \begin{minipage}[b]{0.5\textwidth}
\resizebox{60mm}{!}{\includegraphics[width=1.5in,height=1.5in]{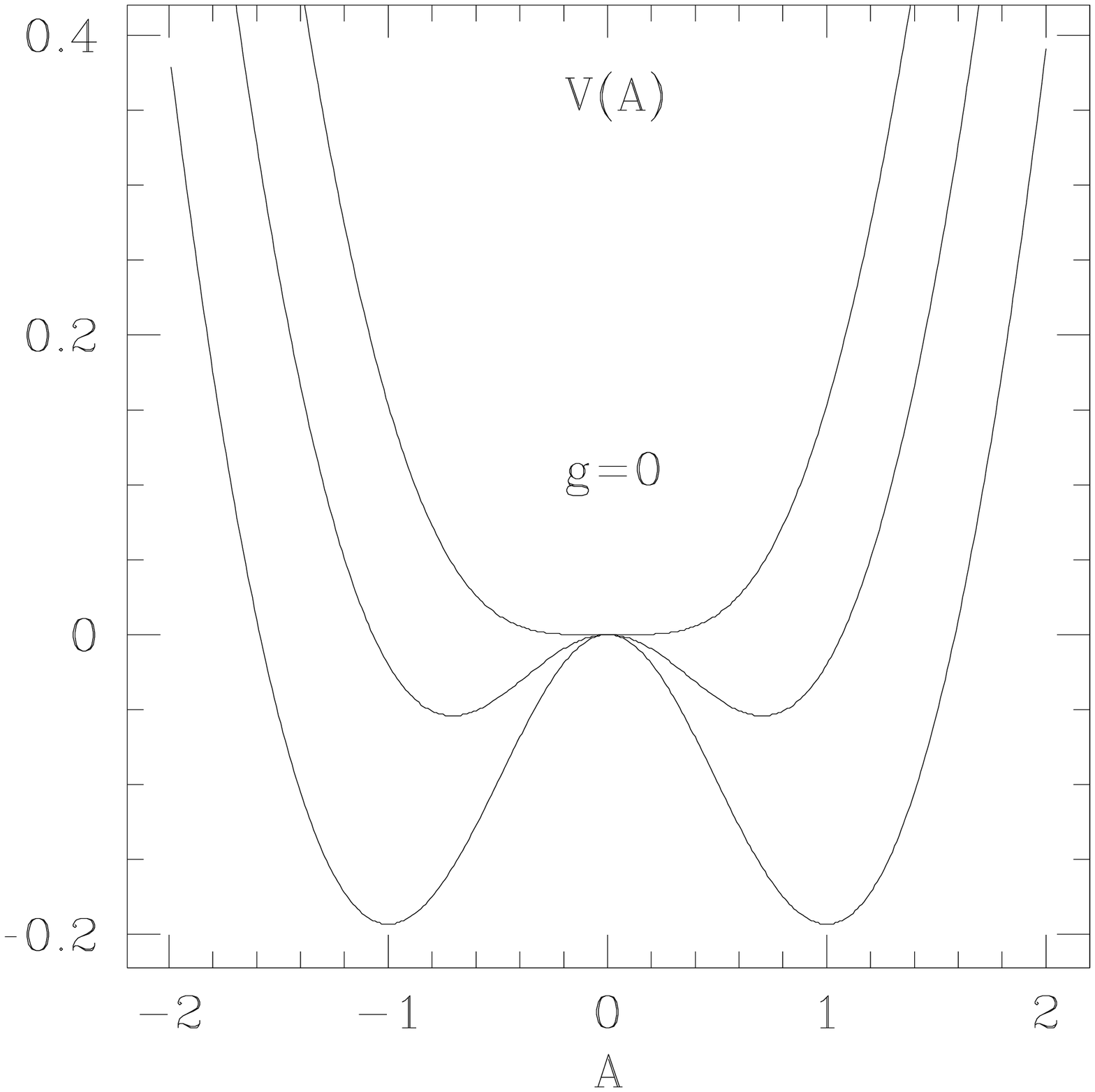}}
 \end{minipage}%
 \begin{minipage}[b]{0.5\textwidth}
\resizebox{60mm}{!}{\includegraphics[width=1.5in,height=1.5in]{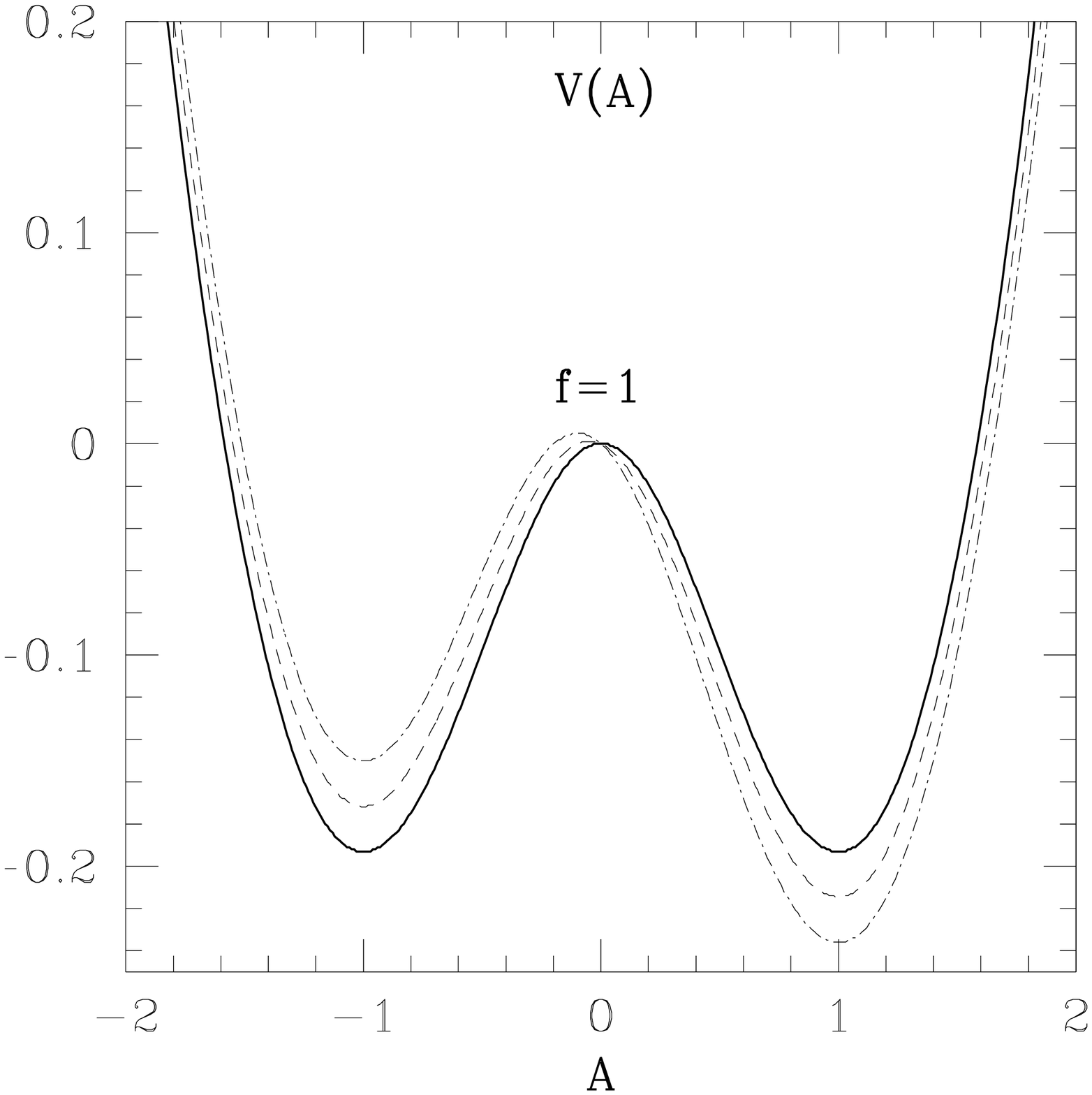}}
 \end{minipage}%
\caption{Left: Potential $V({\cal A})$ for $g=0$ and varying fidelity $f$.
From bottom to top, $f=1,0.75,0.5$. Right: Potential for $f=1$ and varying
asymmetry $g$. The bold line corresponds to the symmetric case, $g=0$. The
dash line to $g=0.1$ and the dot-dash line to $g=0.2$.}
\end{figure}

\subsection{Introducing Spatial Dependence}

In order to introduce spatial dependence for the concentrations, we follow the
usual procedure in the phenomenological treatment of phase transitions
\cite{Gunton, Langerrev}, by rewriting the total time derivatives in eqs.
\ref{SAeqs1} as $d/dt\rightarrow \partial/\partial t - k\nabla^2$, where $k$ is
the diffusion constant. Some illustrative values of $k$ are: 
$k=10^{-9}$m$^2$s$^{-1}$ for molecular diffusion in water and
$k=10^{-5}$m$^2$s$^{-1}$  for air. The diffusion time-scale in a length $L$ is
$\tau_{\rm diff} = L^2/k$.

It is convenient to introduce the dimensionless time and space variables,
$t_0\equiv \lambda_0t$ and $x_0\equiv x\sqrt{\lambda_0/k}$, respectively. The
equations are then solved in terms of the dimensionless variables. Dimensionful
values are obtained for particular choices of the parameters $k_S,~Q$, and $k$.
For example, using as nominal values  $k_S\sim 10^{-25}{\rm cm}^3{\rm
s}^{-1}$,  $Q\sim 10^{15}{\rm cm}^{-3}{\rm s}^{-1}$ and $k$ for water, one
obtains, $t\simeq 2.3\times 10^{-3}t_0$ y and  $x\simeq 8.5 \times10^{-3}x_0$m,
while $\lambda_0\simeq \sqrt{2}\times 10^{-5}$s$^{-1}$.

The main consequence of introducing spatial dependence is that now the net
chiral asymmetry will evolve in both space and time. With a racemic or 
near-racemic initial distribution, a typical spatial volume ${\cal V}\gg
\xi^3$  will coarsen into domains of left and right-handed polymers, separated
by an interface or domain wall with approximate thickness $\sqrt{k/\lambda_0}$:
an initially racemic solution gradually separates into chiral domains.  The
``wall'' between homochiral domains is to be interpreted as the region of space
which remains racemic [${\cal A}({\bf x},t)=0$.] In order for an initially
racemic mix to evolve toward homochirality, the walls separating the left and
right domains must move. We will discuss in detail below what physical processes
may trigger the wall mobility.

In one spatial dimension, and for $f=1$ and $g=0$, the solution is well
approximated by a ``kink'' profile obtained in the static limit ($\dot{\cal
A}=0$) as  ${\cal A}_k(x)=\tanh(-\alpha x\sqrt{\lambda_0/k})$, where $\alpha$
can be found numerically to be $\alpha\simeq 0.58$ \cite{BM}.

Using the dimensionless variables defined above,
the energy of a static spatially-extended configuration in $d$ spatial 
dimensions is given by
$E[{\cal A}] = \left (k/\lambda_0\right )^{d/2}\int d^dx_0
\left [\frac{1}{2}\nabla_0{\cal A}\cdot\nabla_0{\cal A} + V({\cal A})\right ],$
where appropriate boundary conditions must be imposed.

\section{Chiral Selection as a Phase Transition I: No Bias}

Gleiser and Thorarinson coupled the net chirality ${\cal A}({\bf x},t)$ to an
external environment modeled by a stochastic force ($\zeta(t, {\bf x})$) with
zero mean  ($\langle\zeta\rangle=0$) and two-point correlation function
$\langle \zeta({\bf x'},t')\zeta({\bf x},t)\rangle =  a^2\delta({\bf x'}-{\bf
x}) \delta(t'-t)$, where $a^2$ measures the strength of the environmental
influence. For example, in Brownian motion, $a^2=2\gamma k_BT$, where $\gamma$
is a viscosity coefficient. With the dimensionless variables introduced above,
the noise amplitude scales as  $a^2_0\rightarrow
\lambda_0^{-1}(\lambda_0/k)^{d/2}a^2$.  Notice that by writing $\gamma =
\lambda_0\gamma_0$ and identifying the noise amplitudes,
$2(\lambda_0\gamma_0)(k_BT) = \lambda_0 (k/\lambda_0)^{d/2}a_0^2$, we obtain
that  the thermal energy, $k_BT$, has dimensions of $(k/\lambda_0)^{d/2}$ and
thus of [length]$^d$ as it should. The equations of motion in the presence
of noise and no bias are
\begin{eqnarray}
\label{aeq}
\lambda_0^{-1}\left(\frac{\partial{\cal S}}{\partial t} 
-k\nabla^2{\cal S}\right)&=&
1 - S^2 + \lambda_0^{-1}\xi({\bf x},t)\\
\lambda_0^{-1}\left( \frac{\partial{\cal A}}{\partial t}
-k\nabla^2{\cal A}\right)&=&
2f\frac{{\cal S}{\cal A}}{{\cal S}^2+
{\cal A}^2} - {\cal S}{\cal A} + \lambda_0^{-1}\xi({\bf x},t)~.
\end{eqnarray}

\subsection{Critical Point for Chiral Symmetry Breaking}

As shown by GT, in the absence of chiral bias and with $f=1$ and ${\cal S}=1$
(an attractor in phase space), the system described by eq. \ref{aeq} has
critical behavior controlled by the noise amplitude $a^2$. For $a$ above a
critical value, $a_c$, $\langle {\cal A}\rangle \rightarrow 0$ and the chiral
symmetry is restored. The brackets denote spatial averaging, $\langle {\cal
A}\rangle = (1/V)\int {\cal A} d^3x$, where $V$ is the volume.  In analogy with
ferromagnets, where above a critical temperature the net magnetization is zero,
one may say that above $a_c$ the stochastic forcing due to the external
environment overwhelms any local excess of $L$ over $R$ enantiometers:
racemization is achieved at large scales and chiral symmetry is restored
throughout space. Thus, the history of chirality on Earth and on any other
planetary platform is inextricably enmeshed with its early environmental
history. Although in the present work we will only consider the case of 
``gentle'' (diffusive) evolution of the reaction network, it should be noted
that ``violent'' disturbances may greatly affect the results and, thus, the
final net chiral excess. Preliminary results for 2d turbulent stirring have
been recently presented,  showing that indeed it may accelerate the emergence
of a final homochiral  state \cite{BM}. We are presently pursuing an
alternative approach wherein the interactions with the environment are modelled
stochastically (cf. eqs. \ref{aeq} above) and hope to report on our results
shortly.

The equation dictating the evolution of the enantiometric excess ${\cal A}$ was
solved with a finite-difference method in a $1024^2$ grid and a $100^3$ grid
with $\delta t=10^{-3}$ and $\delta x=0.2$,  and periodic boundary conditions.
In 2d, this corresponds to simulating a shallow pool with linear dimensions of
$\ell\sim 200$cm.  In order to obtain the value of $a_c$, the system was
prepared initially in a homochiral phase chosen to be  $\langle {\cal A}\rangle
(t=0)=1$. The equation was then solved for different values of the external
noise amplitude, $a$. As shown in GT, for $a^2>a_c^2\simeq 0.65
(k/\lambda_0)^{3/2}$, $\langle {\cal A}\rangle \rightarrow 0$, that is, the
system becomes racemized. $\langle {\cal A}\rangle$ approaches a constant for
large times, indicating that the reaction network reaches equilibrium with the
environment. For $d=2$, $a^2_c\simeq 1.15 (k/\lambda_0)$. In figure 2
we show the phase diagram for chiral symmetry restoration in 3d.

\begin{figure}
\label{tcrit}
\hspace{1.in}
\includegraphics[width=3.5in,height=3.5in,angle=-90]{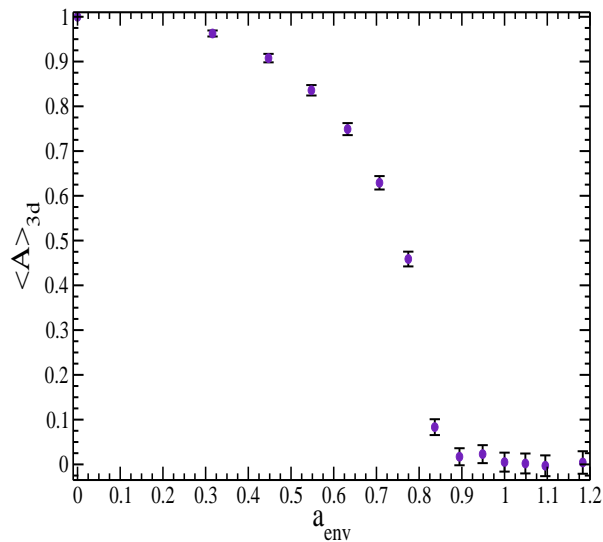}
\caption{Phase diagram for the 3d volume-averaged net chirality $\langle
{\cal A}\rangle_{\rm 3d}$ as a function of external noise
$a_{\rm env}$. The error bars denote ensemble averaging over 20 runs.}
\end{figure}

\subsection{Ginzburg Criterion in the Absence of Chiral Bias}

It is possible to gain much insight into the relationship between the
microscopic and mean-field approaches using the Ginzburg  criterion
\cite{Landau80}. With $g=0$ and $f=1$, the potential $V({\cal A})$ is  a
symmetric double-well with minima at ${\cal A}_{\pm}=\pm 1$ and a  maximum at
${\cal A}=0$. ${\cal A}= {\cal A}_{\pm}$ are fixed points: if
the system is prepared away from them, it will evolve towards them.
When domains of both phases are present, they will be separated by domain
walls. In the absence of bias ($g=0$) and of environmental
coupling ($a=0$),  the only force on the walls comes from surface tension: the
walls will straighten in order to minimize their radii within a given
volume. When the average wall radius, ${\bar R}(t)$, is comparable to the
linear dimensions of the confining volume ($\ell$), the wall will stall or
move exceedingly slowly.  Given that left and right-handed life forms cannot
coexist in the same domain, this simple model predicts, quite reasonably, that
other factors in early-Earth's history have intervened to promote the observed 
homochirality.

Within the continuous description, the smallest volume is the  correlation
volume, $V_{\xi}\simeq 4\xi^3$, where the correlation length $\xi$ is related
to the potential $V({\cal A})$ by  $\xi^{-2} = V''({\cal A}=\pm 1)$
\cite{Landau80}. The energy $E_G$ required to flip a correlation-volume cell of a
given chirality into one of opposite chirality is given by the energy barrier
($\Delta V$) times the correlation volume: $E_G = V_{\xi}\Delta V$, where
$\Delta V = |V(0)-V(\pm 1)|$. On the other hand, if $E_f$ is the energy to flip
the chirality of  a single molecule and $N_{\xi}$ is the average number of
molecules in a correlation volume, then $E_G = N_{\xi}E_f$. Equating both
expressions for $E_G$ we obtain, $E_f = \frac{V_{\xi}}{N_{\xi}}\Delta V.$

From the expression for the potential, eq. \ref{pot},  we obtain, with $f=1$
and $g=0$, $\Delta V=0.193$ and $\xi = (k/\lambda_0)^{1/2}$, so that 
$V_{\xi}\simeq 4(k/\lambda_0)^{3/2}$.  In order to estimate $N_{\xi}$, note
that the parameters $Q$ and $k_S$ define the microscopic length-scale $\xi_{\rm
micro} = (Q/k_S)^{-1/6}$. Using $Q=10^{15}{\rm cm}^{-3}{\rm s}^{-1}$ and
$k_S=10^{-25}{\rm cm}^3{\rm s}^{-1}$, we obtain, $\xi_{\rm micro} \simeq
2.154\times 10^{-7}$cm and  $N_{\xi}\simeq (\xi/\xi_{\rm micro})^3$. Thus,
$E_f\simeq 4\xi^3_{\rm micro} \Delta V \simeq 7.7\times 10^{-21}{\rm cm}^3$.

This energy can be compared with the results from the numerical study of chiral
symmetry breaking in GT, who found $a_c^2\simeq  0.65
(k/\lambda_0)^{3/2}\simeq 0.4{\rm cm}^3$.  This is the critical energy for a
correlation-volume cell to flip chirality. So, we must divide it by the number
of molecules in a correlation volume in order to obtain the critical energy per
molecule,  $E^{\rm num}_c \simeq 6.5\times 10^{-21}{\rm cm}^3$. The ratio of
the  two energies is, $E_f/E^{\rm num}_c \simeq 1.18$, a nice agreement between
the theoretical prediction from the Ginzburg criterion and the numerical
results.

\section{Chiral Selection as a Phase Transition II: Including Bias}

Environmental effects, if above a certain threshold, may destroy any net
chirality, restoring the system to a racemic state.  Once external
perturbations cease, the system will relax to its thermodynamically preferred
state as it evolves toward final equilibrium. Within the present mean-field
model, this evolution is characterized by a competition between left and
right-handed domains separated by interfaces. This evolution is determined by
the initial domain distribution and by the forces acting on the interfaces.

\subsection{Percolation Constraints on Chiral Bias}

One may think of the domains of each chirality as two competing populations
immersed in an  environment at ``temperature'' $T$. We use quotes to stress
that this external influence may be attributed to several different sources of
white noise with Gaussian amplitude $a^2$.  Consider a large volume ${\cal
V}=V_L + V_R\gg V_{\xi}$,  where $V_{L(R)}$ is the total volume in left(right)
domains. If the fractional volumes of {\it both} left ($V_L/{\cal V}$) and
right ($V_R/{\cal V}$) domains  exceed a critical value $p_c$, they both
percolate \cite{percolation} and the total volume will be a convoluted
structure similar to that of a sponge, with regions of left and right chirality
separated by a  thin interface. If, instead, only one of the two  fractional
volumes exceeds $p_c$, the volume will be percolated by the dominant
handedness.  The isolated domains of the opposite handedness will shrink and
disappear.

Let $p_{L(R)}$ be the probability that a randomly chosen correlation-volume
cell will be left(right)-handed. In thermal equilibrium, the relative
probabilities obey, $p_L/p_R = \exp[-\Delta F/k_BT]$, where $\Delta F$ is the
free-energy difference between the two populations. Of course, if they are
equally probable (no bias), $p_L/p_R =1$. Within the mean-field model,
the free-energy difference between the two enantiometric phases is
$\Delta F = \Lambda V_{\xi}~, \Lambda\equiv |V({\cal A}_L) - V({\cal A}_R)|$,
where $V({\cal A}_{L(R)})$ is evaluated at the potential minima. For the
potential of eq. \ref{pot} with $f=1$, $\Lambda = 2g(1-\pi/4)$. 

For temperatures above the Ginzburg temperature ($T_G$),  thermal fluctuations
may drive the domains to flip their chiralities. As discussed in section 4.2,
the associated energy scale is $E_G=k_BT_G\simeq V_{\xi}\Delta V$
\cite{Landau80}.  Given that we will be mostly interested in very small biases
($\Delta V\gg \Lambda$), we may use the expression for $\Delta V$ obtained by
taking $g=0$. Thus, as the temperature drops below $T_G$, the chiral flipping
is exponentially suppressed and the two populations are fixed by,
$\frac{p_L}{p_R}|_{T\leq T_G} = \exp\left
[-\frac{\Lambda}{\Delta V}\right ].$
From the results above, with $f=1$, we obtain $p_L/p_R|_{T_G} =
\exp[-2g(1-\pi/4)/0.193]\simeq \exp[-2.224g]$.

The ratio $p_L/p_R|_{T_G}$ sets the initial conditions for the subsequent
dynamics of the system. The key point is whether one or both phases percolate.
This is decided by comparing the probabilities $p_{L(R)}$  with the critical
percolation probability, $p_c$. Now, $p_c$ depends on dimensionality and
somewhat less on the shape of the lattice cell. As an illustration, and to be
consistent with the numerical simulations in GT which were performed on 3d
cubic lattices, we take $p_c=0.31$, the result for cubic lattices in 3d
\cite{percolation}.  Using that $p_L+p_R=1$, we obtain that for both phases to
percolate, $g\leq g_c=0.36$. This is an upper bound on chiral bias in order for
both types of domains to coexist and percolate throughout the volume. Clearly,
if $g>0.36$ only one domain percolates and chirality is firmly determined.
However, unless some presently unknown effect strongly biases one enantiometer,
$g$ is most probably much smaller and both phases will initially percolate. 

In figure 3 we show an example where both phases percolate, with $g=0$. The
reader can verify that each phase crosses the entire lattice,  while the two
phases are separated by a thin interface or domain wall. This 2d simulation
used a $1024^2$ lattice with periodic boundary conditions. For $g>0.36$, only
one of the two phases would percolate through the lattice: the minority phase
would be constrained to form finite-volume domains that would shrink and
disappear due to surface tension. We discuss wall dynamics next.

\begin{figure}
\label{percolate}
\hspace{0.25 in}
\includegraphics[width=4in,height=4in]{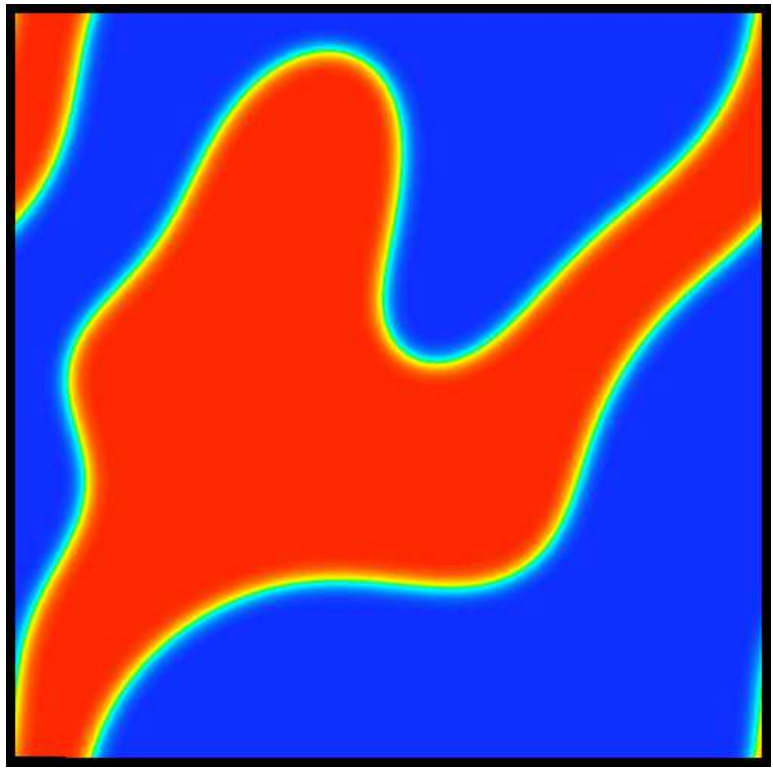}
\caption{Snapshot of evolution for $g=0$, showing the two phases percolating the
entire lattice and separated by a thin domain wall.}
\end{figure}

\subsection{Wall Dynamics in the Presence of Bias}

Once the flipping between phases ceases below $T_G$ and percolation occurs,
the wall network begins to evolve. 
Barring external influences, 
two forces will act on the walls: the curvature pressure $p_s$ will act to
straighten the walls, while the biasing pressure $p_g=\Lambda$ will tend to 
accelerate the walls toward the unfavored phase.

The curvature pressure can be written as
$p_s\simeq \sigma/{\bar R}(t)$, where $\sigma = \sigma_0(k/\lambda_0)^{1/2}$
is the surface tension, with 
$\sigma_0 = \int dx_0 \left [\frac{1}{2}(\nabla_0 A_k)^2 + V(A_k) \right ]= 
\int_{-1}^{+1}dA [2V(A)]^{1/2}$,
and ${\bar R}(t)$ is its average radius. 
Using eq. \ref{pot} with $f=1$ and $g=0$, we obtain numerically that 
$\sigma_0= 0.789$. [Corrections from a small $g$ are quite negligible.]

At $T_G$, both the average wall curvature and separation will be of order
of the correlation length $\xi=(k/\lambda_0)^{1/2}$. Thus, we can write
\begin{equation}
\label{initialcond}
\frac{p_g}{p_s}|_{T_G} = \frac{2g(1-\pi/4)}{\sigma_0} \simeq 0.544g~.
\end{equation}
Unless $g$ is unrealistically large, the motion is initially dominated by the
surface pressure and walls will tend to straighten before the biasing pressure
becomes effective. There are thus two possibilities: given a large confining
volume with linear length-scale $\ell$, either $p_g$ becomes active
before the walls straighten to $\ell$  (${\bar R} \rightarrow \ell$) or
after. In both cases, once $p_g$
does become active, the walls will move toward the unfavored phase so that the
volume $\sim \ell^3$ will eventually become homochiral. 

The key question is thus whether this converting mechanism has enough time to
take place in early-Earth given what we  know of its prebiotic history and the
magnitude of biasing sources proposed so far. In order to answer this question,
we write the average wall radius as ${\bar R}(t) = {\bar
R}_0(t)[k/\lambda_0]^{1/2}$ so that $p_g/p_s (t)= 0.544g{\bar R}_0(t)$, with
${\bar R}_0(t=t_G=0) = 1$. In the last expression we set the time to zero at
wall formation time. Thus, for the biasing pressure to dominate,
the average wall radius must grow to satisfy ${\bar R}_c  \geq
[0.544g]^{-1}$:  small values of $g$ imply in very large radii. We must
next estimate the time ($t_g$) it takes for this critical value to be achieved.
The equation controlling the wall radius is
\begin{equation}
\label{curvaturegrowth}
\frac{1}{\lambda_0}\frac{d{\bar R}}{dt} = \sigma \left (\frac{\xi}{{\bar R}(t)}
\right )~.
\end{equation}
The solution with ${\bar R}_0(0)=1$ is
\begin{equation}
\label{solutionradius} 
{\bar R}(t) = \left (k/\lambda_0\right )^{1/2}\left [1 + 2\sigma_0t_0
\right ]^{1/2}.
\end{equation}
Note that at formation time ($t_0=0$), the initial radius of the walls is
equivalent to the correlation length and thus to the initial domain size as, 
${\bar R}(0)=\xi=(k/\lambda_0)^{1/2}$. One can see how the initial
length-scale in the problem depends on the two physical parameters, the
diffusion coefficient $k$ and the inverse time-scale $\lambda_0$. 

Substituting the critical value for the radius obtained above and solving
for time,
\begin{equation}
\label{time0bias}
t_{0g} = \frac{1}{2\sigma_0}
\left [\left (\frac{p_s}{p_g}|_{T_G}\right )^2-1\right ]~.
\end{equation}
Using the same values as before for $Q$, $k_S$ and diffusivity
in water, we obtain (taking $g\ll 1$),
$t_g \simeq 4.8\times 10^{-3}g^{-2}{\rm y}~.$ [From here on, y is short for 
year, My(By) for millions(billions) of years, 
and Bya for billions of years ago.]

If we want life's early chirality to be decided, say, within 100 million years
-- sometime between the culmination of heavy bombardment period about 4Bya
\cite{bombardment} and first life about 3.5 Bya \cite{firstlife}, we obtain a
lower bound on the biasing,  $g\geq 7\times 10^{-6}(100 {\rm My}/t_g)^{1/2}$.
Values of $g$ obtained from WNC are too small to promote homochirality on early
Earth. The situation with CPL is not as clearcut, but it seems unlikely that
such values of $g$ can be sustained long enough due to the variability of
possible astrophysical sources and the destruction of material by unpolarized
UV light \cite{CPL2,Bonner} which might pass unfiltered through Earth's 
prebiotic atmosphere. A more detailed quantitative study is needed.

For $g\leq 10^{-6}$ it would take longer than the
age of the Univere ($\simeq 14$By) before the biasing becomes active.
Using $t_g$ in the expression for ${\bar R}(t)$ above
gives an estimate for the average radius of the interface by
the time biasing pressure takes over the dynamics:
${\bar R}(t_g)\simeq (k/\lambda_0)^{1/2}(p_s/p_g)|_{T_G}\simeq 2g^{-1}$cm. 

Once biasing pressure takes over, the interface will move toward the unfavored
phase. We can estimate its velocity by studying the equation  governing its
motion \cite{Langerrev}.  Starting with the equation describing the
spatiotemporal evolution of the chiral asymmetry, we look for a solution
moving with constant velocity $v$, $\tilde{\cal A}(x-vt)$. The equation
becomes,
\begin{equation}
\label{wavefront}
\frac{k}{\lambda_0}\frac{\partial^2\tilde{\cal A}}{\partial x^2}
-\frac{\partial V}{\partial \tilde{\cal A}} = - \frac{v}{\lambda_0}
\frac{\partial \tilde{\cal A}}{\partial x},
\end{equation}
where we assumed that the interface can be approximated by the propagation of a
one dimensional front. This is justified considering that the thickness of the
interface is of order $\xi$ and thus {\it much} smaller than its average
radius at $t> t_g$. Note also that upon switching space for time, the
equation describes a particle moving in the presence of a velocity-dependent
viscous force in a potential $-V({\cal A})$. The solution satisfying the
asymptotic boundary condition ${\cal A} (x\rightarrow \pm\infty)=\pm 1$ is
determined by the interface's velocity $v$. Integrating eq. \ref{wavefront} by
parts and using that the surface tension $\sigma = \int dx (\partial {\cal
A}/{\partial x})^2$ for $g\ll 1$, we obtain,
\begin{equation}
\label{wallvelocity}
v = \frac{\Delta V}{\sigma_0}\lambda_0\left (\frac{k}{\lambda_0}\right )^{1/2}
\simeq 0.544g\lambda_0\left (\frac{k}{\lambda_0}\right )^{1/2}~,
\end{equation}
where in the last identity we used the values for the potential of eq.
\ref{pot}.
With the same fiducial values for $k$ and $\lambda_0$ as before,  $v \simeq
2.5g$my$^{-1}$. As an illustration, for the wall to convert a distance of
$1$km in $100$My, $g\geq 4\times 10^{-6}$. For the walls to sweep a distance
equivalent to Earth's radius, $g\geq 2.6\times 10^{-2}$. A small bias doubly
compromises the conversion of racemic prebiotic chemistry to homochirality: i)
the time for the bias to take over, $t_g\simeq 4.8\times 10^{-3}g^{-2}$y can
easily exceed the age of the Universe even for values of the bias much larger
than the ones proposed thus far; and ii) once the bias takes over, the distance
swept by the wall, $d_{\rm wall}(t)\simeq 2.5g(t/{\rm y})$m, can be
exceedingly small even for large times of order $100$My.

\subsection{Nucleation-Induced Homochirality in Prebiotic Environments}

Given that for small bias the wall motion will not set in for a very long time,
it is legitimate to ask whether the conversion to the final homochiral phase
may happen via homogeneous nucleation \cite{Langer67, Gunton}. In this case, a
nucleus of the favored phase will thermally nucleate within the unfavored phase
with a  rate per unit volume $\Gamma(T,g)$ controlled by the Arrhenius factor
\begin{equation} 
\Gamma \simeq \lambda_0(\lambda_0/k)^{3/2}\exp[-E_g({\cal A}_b)/k_BT]~,
\end{equation}
where
$E_g({\cal A}_b)$ is the energy  of the so-called ``bounce'' or critical
nucleus. In the thin-wall approximation, valid in the limit of small asymmetry
between the two cases and thus here ($\Lambda/\Delta V \ll 1$),  the energy of
the bounce of radius $R$ is well-approximated by 
$E(R) = -\frac{4\pi}{3}R^3\Delta V +4\pi R^2\sigma$.
Extremizing this expression, we find, for the radius of the critical nucleus
and its associated energy,
$R_c=2\sigma/\Delta V\simeq 1.84 g^{-1}(k/\lambda_0)^{1/2}$ and
$E(R_c)=\frac{16\pi}{3}\frac{\sigma^3}{(\Delta V)^2}\simeq 44.5g^{-2}
(k/\lambda_0)^{3/2}$,
where in the last expressions we used the potential of eq. \ref{pot}.  Note how
$E(R_c) \propto g^{-2}$. From the expression for the nucleation rate we can
estimate the time-scale for nucleation in a given prebiotic volume $V_{\rm pb}$
as 
\begin{equation}
\label{tnuc}
\tau_{\rm nuc} \simeq (\Gamma V_{\rm pb})^{-1} \simeq 1.3\times 10^{-15}\left
(\frac{10^6{\rm m}^3}{V_{\rm pb}}\right )\exp[89/g^2a^2]{\rm y}~,
\end{equation}
where the last expression was obtained using the usual fiducial values
for $k$ and $\lambda_0$ and the relation $a^2=2k_BT$ for the environmental
fluctuation-inducing noise amplitude.

Given that the nucleation time-scale decreases with volume, as an illustration 
let us consider a large volume of the unbiased phase, say a ``shallow'' 
cylindrical pool with volume $V_{\rm pb} =\pi (10^3 {\rm m})^2(100 {\rm m})=
\pi\times 10^8 {\rm m}^3$. Within this volume, $\tau_{\rm nuc} \simeq 4\times
10^{-18}\exp[89/g^2a^2] {\rm y}$. If we impose, realistically, that $\tau_{\rm
nuc}\leq 100$My, we obtain a bound on the critical nucleation barrier
$E(R_c)/k_BT\leq 58.5$. For small asymmetries, this bound cannot be satisfied.
Indeed,  in terms of the asymmetry $g$, and using that a typical noise
amplitude capable of inducing sizeable fluctuations is $a^2\simeq 0.5$ (cf.
GT), we obtain $g \geq 1.74$, an unrealistically large value. For a volume
with extension comparable to the Earth's radius, $V_{\rm pb}=\pi (6.5\times
10^6 {\rm m})^2(100 {\rm m})=1.3\times 10^{16} {\rm m}^3$, the bound becomes
$g\geq 1.53$. We conclude that homogeneous nucleation cannot resolve the
chirality issue. It remains to be seen if environmental disturbances may
promote a faster nucleation rate \cite{GH}.

\section{Summary and Outlook}

The spatiotemporal evolution of prebiotic homochirality was investigated in the
context of an autocatalytic reaction network featuring enantiometric
cross-inhibition catalysed by dimers and chiral bias. Domains of opposite
chirality, separated by thin interfaces, compete for dominance. It was shown
that the dynamics of the domain network is determined by the percolation
properties of its initial distribution and subsequently by the two main forces
acting on the interfaces, surface tension and chiral bias. Small biases of
$g\leq 10^{-6}$ were shown to be inefficient to drive  reasonably-sized reactor
pools toward homochirality within presently-accepted time-scales for the origin
of life on Earth.  As a consequence, the present calculations indicate that
WNCs cannot explain the observed homochirality of life's  biomolecules. CPL
remains a remote possibility, albeit current sources do not look promising
either in magnitude and duration. Also, it should be noted that unpolarized UV
may destroy any early enantiometric excess.

The results obtained assume that the polymerization dynamics can be captured by
truncating the reaction network to $n\leq 2$ and that the dynamics of dimers is
enslaved by that of monomers (the adiabatic approximation). These
approximations imply that complete chiral separation can only occur with
perfect fidelity $f=1$. Going beyond involves solving the complete network of
spatiotemporal rate equations for larger values of $n$ and for varying
fidelity, a computer-intensive, but not impossible, task. However,  given that
the formation rate of higher $n$ polymers will necessarily be slower, we
believe that the results obtained here capture at least qualitatively the
essentials of the more general case. It remains a challenging open question
whether a simple transformation could be found to reduce the biased higher-$n$
system to an effective field theory as done here for $n=2$.

What other possible sources of bias could have driven Earth's prebiotic
chemistry toward homochirality? We cannot rule out the possibility that some
unknown chemical bias satisfying the above bound might have been  active.
Another, highly unnatractive, possibility is that an unlikely large statistical
fluctuation towards one  enantiometer did occur and established the correct
initial conditions.  Possible bombardment from meteors contaminated with chiral
compounds could also have jump-started the process, although one still needs to
explain how the chiral excess formed in the meteors in the first place. 

It seems to us that the answer to this enigma will be found in the coupling of
the reaction network to the environment. We note again that the results
obtained here are within the diffusive, and hence ``gentle,'' evolution towards
homochirality. Early-Earth, however, was a dramatic environment. Given the
nonlinear properties of the spatiotemporal equations describing the evolution
towards homochirality, environmental disturbances, such as meteoritic impacts
or volcanic eruptions, must have played a key role in early-Earth's prebiotic
chemistry.  These disturbances, if violent enough, would certainly affect the
evolution of the chiral domain network and possibly change the bounds obtained
in the present work. Gleiser and Thorarinson proposed to model the coupling to
an external disturbance stochastically \cite{GT}. Within their framework,
results will depend on how the amplitude of the external ``noise'' compares
with the critical value described in section 4.1. Preliminary results indicate
that large enough noises (modelling external influences) may redirect the
direction of homochirality entirely, erasing any previous evolution toward
either handedness. Further work along these lines is underway. In a different
approach, Brandenburg and Multam\"aki suggested that hydrodynamic turbulence
could have sped up the march toward homochirality \cite{BM}. In either case, it
is clear that the evolution toward homochirality, as that of life itself,
cannot be separated from Earth's early environmental history. 

The author thanks Gustav Arrhenius, Jeffrey Bada, Freeman Dyson, Leslie Orgel,
and Joel Thorarinson for stimulating discussions. He also thanks Joel
Thorarinson for producing figure 3.


\end{document}